\documentstyle [12pt,aaspp4]{article}
\def\beq{\begin{equation}}
\def\eeq{\end{equation}}
\def\bey{\begin{eqnarray}}
\def\eey{\end{eqnarray}}
\def\pppm{\rm P^3M}
\def\mpc{\,h^{-1}{\rm {Mpc}}}
\def\kms{\,{\rm {km\, s^{-1}}}}

\def\br#1{{\mathbf r}_{#1}}
\def\bs#1{{\mathbf s}_{#1}}
\def\zetarr{\zeta(r_{12},r_{23},r_{31})}
\def\zetass{\zeta(s_{12},s_{23},s_{31})}
\def\scycl{(s_{12},s_{23},s_{31})}
\def\rpcycl{(r_{p12},r_{p23},r_{p31})}
\def\rppicycl{(r_{p12},r_{p23},r_{p31},\pi_{12},\pi_{13})}
\def\zetazrprp{\zeta_z(r_{p12},r_{p23},r_{p31},\pi_{12},\pi_{13})}
\def\zetaru{\zeta(r,u,v)}
\def\zetasu{\zeta(s,u,v)}

\def\Qru{Q(r,u,v)}
\def\Qsu{Q_{red}(s,u,v)}

\def\Qrpu{Q_{proj}(r_{p},u,v)}

\def\Pirpu{\Pi(r_p,u,v)}
\def\nbar#1{{\bar n}({\mathbf r}_{#1})}
\def\nbas#1{{\bar n}({\mathbf s}_{#1})}
\def\xiz#1{\xi_z(r_{p#1},\pi_{#1})}
\def\xir#1{\xi(r_{#1})}
\def\xis#1{\xi(s_{#1})}
\def\wrp#1{w(r_{p#1})}
\def\gs{\mathrel{\raise1.16pt\hbox{$>$}\kern-7.0pt
\lower3.06pt\hbox{{$\scriptstyle \sim$}}}}
\def\ls{\mathrel{\raise1.16pt\hbox{$<$}\kern-7.0pt
\lower3.06pt\hbox{{$\scriptstyle \sim$}}}}
\def\gtsima{$\; \buildrel > \over \sim \;$}
\def\ltsima{$\; \buildrel < \over \sim \;$}
\def\prosima{$\; \buildrel \propto \over \sim \;$}
\def\gsim{\lower.5ex\hbox{\gtsima}}
\def\lsim{\lower.5ex\hbox{\ltsima}}
\def\simgt{\lower.5ex\hbox{\gtsima}}
\def\simlt{\lower.5ex\hbox{\ltsima}}
\def\simpr{\lower.5ex\hbox{\prosima}}

\begin{document}
\title {
The Three-point Correlation Function of Galaxies Determined
from the Las Campanas Redshift Survey}
\author {Y.P. Jing$^{1}$, G. B\"orner$^{1,2}$} 
\affil{$ ^1$Research Center for the Early Universe,
School of Science, University of Tokyo, Bunkyo-ku, Tokyo 113, Japan}
\affil {$ ^2$Max-Planck-Institut f\"ur Astrophysik,
Karl-Schwarzschild-Strasse 1, 85748 Garching, Germany}
\affil {e-mail: jing@utaphp2.phys.s.u-tokyo.ac.jp, ~ grb@mpa-garching.mpg.de}
\received{---------------}
\accepted{---------------}

\begin{abstract}
  We report the measurement of the three-point correlation function
  (3PCF) of galaxies for the Las Campanas Redshift Survey (LCRS). We
  have not only measured the 3PCF in redshift space but also developed
  a method to measure the projected 3PCF which has simple relations to
  the real space 3PCF.  Both quantities have been measured as a
  function of triangle size and shape with only a fractional
  uncertainty in each individual bin. Various tests derived from mock
  catalogs have been carried out to assure that the measurement is
  stable and that the errors are estimated reliably.  Our results
  indicate that the 3PCFs both in redshift space and in real space
  have small but significant deviations from the well-known
  hierarchical form.  The 3PCF in redshift space can be fitted by
  $\Qsu=0.5\cdot 10^{[0.2+0.1({s\over s+1})^2]v^2}$ for $0.8<s_{12}<8
  \mpc$ and $s_{31}<16\mpc$, and the projected 3PCF by
  $\Qrpu=0.7r_p^{-0.3}$ for $0.2<r_{p12}<3\mpc$ and $r_{p31}<6\mpc$
  ($s$ and $r_p$ are in unit of $\mpc$), though a systematic weak
  increase of $\Qrpu$ with $v$ at $r_p>1\mpc$ is noted.  The
  real-space $\Qru$ for $0.2 \ls r_{12} \ls 3\mpc$ and $r_{31}\ls
  6\mpc$ can be well described by {\it half} the mean 3PCF predicted
  by a CDM model with $\Omega_0 h=0.2$.

  The general dependence of the 3PCF on triangle shape and size is in
  qualitative agreement with the CDM cosmogonic models. Quantitatively
  the 3PCF of the models may depend on the biasing parameter and the
  shape of the power spectrum, in addition to other model
  parameters. Taking our result together with the constraints imposed
  by the two-point correlation function and the pairwise velocity
  dispersion of galaxies also obtained from the LCRS, we find that we
  have difficulties to produce a {\it simple} model that meets all
  constraints perfectly.  Among the CDM models considered, a flat
  model with $\Omega = 0.2$ meets the 2PCF and PVD constraints, but
  gives higher values for the 3PCF than observed.  This may indicate
  that more sophisticated bias models or a more sophisticated
  combination of model parameters must be considered.

\end{abstract}


\section {Introduction}
Correlation functions are very powerful statistics to describe the
large scale structures in the Universe (\cite{p80} 1980, hereafter
P80). The lowest order, the two-point correlation function (2PCF)
$\xi(r)$, has been widely used to measure the clustering strength of
galaxies and to confront models of cosmic structure formation. Quite a
number of large galaxy catalogs, both angular and redshift, have been
used to determine the 2PCF, and this statistic has now been
established quite well (\cite{jmb98} 1998, hereafter JMB98;
\cite{letal98} 1998 in preparation; \cite{letal97} 1997;
\cite{tetal97} 1997; \cite{rspf97} 1997; \cite{baugh96} 1996;
\cite{hetal96} 1996; \cite{sw95} 1995 for a review before 1995). This
statistic has produced several constraints on theoretical models
already despite the fact that there are many ingredients to a specific
model which can be optimally adapted to the properties of the galaxy
sample. The cosmological parameters, like the initial power spectrum
of the DM component and the bias, i.e.  the difference in the
clustering of galaxies and DM particles, can all be adjusted to some
extent.

The three-point correlation function (3PCF) $\zetarr$ is a further
statistic useful in characterizing the clustering of galaxies
(P80). Its measurement can give additional constraints
 for cosmogonic models. The determination of the
3PCF was pioneered by Peebles and his coworkers in the seventies. Based on
their careful analysis of the Lick and Zwicky angular catalogs of
galaxies they propose a so-called ``hierarchical" form
\beq\label{hier} 
\zetarr=Q\Bigl[\xir{12}\xir{23} +\xir{23}\xir{31}
+\xir{31}\xir{12}\Bigr] 
\eeq 
with the constant $Q\approx 1.29\pm 0.2$. This form is valid for
scales $r\ls 3\mpc$ (P80). The analysis of the ESO-Uppsala catalog of
galaxies (\cite{lau82} 1982) by \cite{jmb91} (1991) supports this
result.  The 3PCF was also examined for the CfA, AAT and KOSS redshift
samples of galaxies (\cite{p81} 1981; \cite{betal83} 1983; \cite{ej84}
1984; \cite{hfms89} 1989). Because all these redshift samples are
small (with $<2000$ galaxies), these authors were not able to examine
the validity of the hierarchical form in redshift space. Instead they
just forced a fit of the hierarchical form and obtained the value of
$Q$. The $Q$ value of redshift samples obtained in this way is around
$0.6$ (\cite{ej84} 1984), much smaller than the value advocated by
Peebles and his coworkers. The difference may partially be attributed
to the redshift distortion effect which reduces the $Q$ value
(\cite{m94} 1994).  The skewness analysis of the 1.2 Jy IRAS survey of
galaxies has given a similar $Q$ value (\cite{betal93} 1993).
 
The hierarchical form (Eq.~\ref{hier}) is purely empirical. There is
no solid theoretical argument supporting this form. In contrast, the
second-order perturbation theory predicts that $Q$ depends on the
shape of the triangle and on the slope of the linear power spectrum
(\cite{f84} 1984) in the linear regime.  For a CDM-like power spectrum
with a slope  which changes with the scale, $Q$ then varies with the size
and shape of a triangle (\cite{jb97} 1997). Even in the strongly
non-linear regime where the hierarchical form was established, the
CDM models do not seem to obey this form
as demonstrated by \cite{ms94} (1994) based on N-body
simulations. Recently, \cite{yg97} (1997) have re-examined, based on
the BBGKY equation, the stable clustering (strongly non-linear) problem
and pointed out that the hierarchical form holds if the clustering is
stable. The question is if the condition of stable clustering can be
achieved in the real Universe (\cite{jain97} 1997).

The 3PCF of galaxies carries much useful information which is
important for cosmogonic models. The theories based on CDM models
predict that the 3PCF of galaxies depends on the shape of the linear
power spectrum (\cite{f84} 1984; \cite{jb97} 1997) and the galaxy
biasing relative to the underlying mass (\cite{detal85} 1985;
\cite{gf94} 1994; \cite{mjw97} 1997; \cite{mvh97} 1997; \cite{cat98}
1998). It might also be sensitive to a possible non-Gaussianity of the
initial density fluctuation (\cite{fs94} 1994). Furthermore the 3PCF
must be determined accurately if one wants to use the cosmic virial
theorem to obtain the mean density of the Universe.

We have carried out a detailed analysis of the Las Campanas Redshift
Survey (\cite{setal96} 1996), and in this paper we report the
measurement of the 3PCF of galaxies in this survey. We have not only
measured the 3PCF in redshift space but also developed a method to
measure the projected 3PCF which has simple relations to the real
space 3PCF (\S 3.1). Our methods are checked very carefully with the
help of mock catalogs generated from N-body simulations, and the
physical meaning of these two quantities is also investigated (\S
3.2). Our statistical results are compared with  previous work,
with the emphasis on a critical examination of the hypothesis of the            hierarchical form  (\S
3.4). As we will see, the hierarchical form does not seem to be  a good
prescription even in the strong clustering regime. We will present a
new fitting formula in \S 3.3. Implications for cosmogonic models are
discussed in \S 4.

\section{Observational sample and mock catalogs}
The sample used for our analysis is the Las Campanas Redshift Survey
(\cite{setal96} 1996; hereafter LCRS). This is the largest redshift survey,
which is now publicly available. Our sample consists of all galaxies
with recession velocities between 10,000 and 45,000 $\kms$ and with
absolute magnitudes (in the LCRS hybrid R band) between $-18.0$ and
$-23.0$. There are 19558 galaxies in this sample, of which 9480 are in
the three north slices and the rest in the three south slices.  The
survey is a well-calibrated sample of galaxies, ideally suited for
statistical studies of large-scale structure. All known systematic
effects in the survey are well quantified and documented (Shectman et
al. 1996; Lin et al. 1996), and so most can be corrected easily in
statistical analyses. The only exception is the `fiber collision'
limitation which prevents two galaxies in one $\sim 1.5\times 1.5\,
{\rm deg}^2$ field from being observed when they are closer than
$55''$ on the sky, because it is impossible to put fibers on both
objects simultaneously.  Here we will use extensively mock catalogs
generated from N-body simulations to quantify this effect.

The real-space 2PCF and the Pairwise Velocity Dispersion (PVD) have
been determined for the LCRS by JMB98. The redshift-space 2PCF and
power spectrum for this sample were presented by \cite{tetal97} (1997)
and \cite{letal97} (1997) respectively. All these studies have shown
that the LCRS is large enough to accurately measure these low order
statistical quantities. In particular, JMB98 have carried out a
detailed comparison between the observed 2PCF and PVD and the
predictions of currently favoured CDM cosmogonies.  They have used a
large set of mock samples to adequately compare models and
observations.  The construction of mock catalogues from the
simulations, i.e.  photometric catalogues subject to the same
selection effects as the real observations are a very important aspect
of their analysis, because only in this way could the statistical
significance of the results be asserted.  Three spatially flat models
have been considered in JMB98, with ($\Omega_0$, $\lambda_0$,
$\Gamma$,$\sigma_8$)=(0.2,0.8,0.2,1.), (0.3,0.7,0.2,1.), and (1.0,
0.0, 0.5, 0.62), where $\Omega_0$ is the density parameter,
$\lambda_0$ is the cosmological constant, $\Gamma=\Omega_0 h$ and
$\sigma_8$ are the shape parameter and normalization of the CDM power
spectrum (\cite{bbks86} 1986).  All of the models give a steeper 2PCF,
and a higher PVD on small scales than the data. Thus unless galaxies
are biased with respect to the mass with a scale-dependent bias, all
these models can be ruled out.  Unfortunately physical models for a
density or a (not so wanted, but perhaps unavoidable) velocity bias are
not on firm grounds. Therefore in JMB98 a simple, but plausible
phenomenological model for the bias has been suggested. To suppress
the number of pairs in the DM distribution at small separations, it is
assumed that the number of galaxies per unit dark matter mass $N/M$ is
smaller in massive halos than in less massive ones. If a behaviour
such as $N/M\propto M_{cl}^\alpha$ with $\alpha=-0.08$ is used for
clusters of mass $M_{cl}$, the predictions of some CDM models are
consistent with the observational results. The best agreement was
achieved for the flat $\Omega_0=0.2$ model.

We will use 10 mock catalogs of this model to test our
statistical methods and quantify the `fiber collision' effect. Since
this model has reproduced the LCRS 2PCF and PVD, we believe these mock
catalogs are very suitable for this purpose. We shall also use 
these mock samples for model testing, as an example to illustrate the
power of the three-point correlation function in discriminating between models
which have similar two-point correlations. Since the model is a
typical CDM model, we will generalize the discussion to other CDM
models.

\section{The three point correlation function}
\subsection{Definitions and statistical methods}
The three-point correlation function (3PCF) $\zetarr$ is defined,
through the joint probability $dP_{123}$ of finding one object
simultaneously in each of the three volume elements $d\br{1}$,
$d\br{2}$ and $d\br{3}$ at positions $\br{1}$, $\br{2}$ and $\br{3}$
respectively, as follows (P80):
\beq\label{3pcfdf}
dP_{123}=\nbar{1}\nbar{2}\nbar{3}\bigl[1+\xir{12}+\xir{23}+\xir{31}+
\zetarr\bigr] d\br{1} d\br{2} d\br{3} 
\eeq 
where $r_{ij}=|\br{i}-\br{j}|$, and $\nbar{i}$ is the mean density of
galaxies at $\br{i}$. This definition can be applied straightforwardly
to redshift surveys of galaxies to measure the 3PCF $\zetass$ of
galaxies in redshift space (at this point we neglect the anisotropy
induced by the redshift distortion which will be considered
later). Here and below we use $\mathbf r$ to denote the real space and
$\mathbf s$ the redshift space.

The 3PCF of galaxies can be measured from the counts of different
triplets (P80). For this purpose, a sample of randomly distributed
points, which has exactly the same boundaries and the same
observational selection effects as the real survey, is generated. Four
types of distinct triplets with triangles in the range ($s_{12}\pm 1/2 \Delta
s_{12}$, $s_{23}\pm 1/2 \Delta s_{23}$, and $s_{31}\pm 1/2 \Delta
s_{31}$) are counted: the count $DDD\scycl$ of triplets formed by
three galaxies; the count $DDR\scycl$ of triplets formed by two
galaxies and one random point; the count $DRR\scycl$ of triplets
formed by one galaxy and two random points; the count $RRR\scycl$ of
triplets formed by three random points.  Following the definition
[eq(\ref{3pcfdf})], we shall use the following estimator
\bey\label{3pcfred} 
\zetass&=&{27 RRR^2\scycl \times DDD\scycl\over
DRR^3\scycl}\nonumber\\ 
&&- {9 RRR\scycl \times DDR\scycl\over DRR^2\scycl} +2 
\eey
to measure the 3PCF of the galaxies in redshift space. The above
formula is slightly different from the estimator used by Groth \&
Peebles (1977). Here we have extended the argument of \cite{h93} (1993)
for the 2PCF to the case of the 3PCF. The coefficients $27$ and $9$
are due to the fact that only {\it distinct} triplets are counted in
this paper. Since the early work of Peebles and coworkers (P80)
indicates that the 3PCF of galaxies is approximately hierarchical, it
is convenient to express the 3PCF in a normalized form
$Q_{red}\scycl$:
\beq\label{qred} 
Q_{red}\scycl ={\zetass \over \xis{12}\xis{23}
+\xis{23}\xis{31} +\xis{31}\xis{12}} \,.  
\eeq 
It is also convenient to use the variables introduced by Peebles (P80)
to describe the shape of the triangles formed by the galaxy
triplets. For a triangle with the three sides $s_{12}\le s_{23} \le
s_{31}$, $s$, $u$, and $v$ are defined as:
\beq\label{ruv}
s = s_{12}, \hskip1cm u={{s_{23}}\over{s_{\rm 12}}},\hskip1cm
v={{s_{31}-s_{23}}\over{s_{12}}}\,.
\eeq
Clearly, $u$ and $v$ characterize the shape and $s$ the size of a
triangle. We take equal logarithmic bins for $s$ and $u$ with the bin
intervals $\Delta \lg s=\Delta \lg u=0.2$, and equal linear bins for
$v$ with $\Delta v=0.2$. For our analysis, we take the following
ranges for $s$, $u$ and $v$: $0.63\le s\le 10\mpc$ ($6$ bins); $1\le
u\le4$ (3 bins); and $0\le v\le 1$ (5 bins).

A sample of 25,000 random points is first generated. The counts $RRR$
are less than $\sim 5$ for small triangles ($s< 1\mpc$). In order to
suppress the fluctuation induced by the random samples, we have
recalculated the counts $RRR$ for $s_{31}\le 4\mpc$ by generating a
random sample 10 times larger, which ensure that the counts $RRR$ are
at least $\sim 300$ of the interested triangle configurations. We
scaled these counts to 25,000 random points and also use these counts
to get $DRR$ on the small scales since $RRR/DRR$ is constant. However,
it is not trivial to search triplets for so many points.  We have
generalized the ordinary linked-list technique of $\pppm$ simulations
(\cite{he80} 1980) to spherical coordinates to count the triplets. The
linked-list cells are specified by the spherical coordinates, i.e. the
right accession $\alpha$, the declination $\delta$ and the distance
$s$.  With this short-range searching technique, we can avoid the
triplets out of the range specified thus making counting triplets very
efficient.

The 3PCF in  redshift space $\Qsu$ depends both on the real space
distribution of galaxies and on their peculiar motions. Although this
information contained in $\Qsu$ is also useful for the study of the
large scale structures (see \S 4), it is apparent that $\Qsu$ is
different from  $\Qru$ in real space. In analogy with the analysis
for the two-point correlation function, we have determined the
projected three-point correlation function $\Pi\rpcycl$. We
define the redshift space three-point correlation function 
$\zetazrprp$ through:
\bey\label{3pcfzdf}
dP^z_{123}&=&\nbas{1}\nbas{2}\nbas{3}
\bigl[1+\xiz{12}+\xiz{23}+\xiz{31}\nonumber\\
&&+
\zetazrprp\bigr] d\bs{1} d\bs{2} d\bs{3} 
\eey 
where $dP^z_{123}$ is the joint probability of finding one object
simultaneously in each of the three volume elements $d\bs{1}$,
$d\bs{2}$ and $d\bs{3}$ at positions $\bs{1}$, $\bs{2}$ and $\bs{3}$;
$\xiz{}$ is the redshift space two-point correlation function;
$r_{pij}$ and $\pi_{ij}$ are the separations of objects $i$ and $j$
perpendicular to and along the line-of-sight respectively.
The projected 3PCF $\Pi\rpcycl$ is then defined as:
\beq\label{3pcfproj1}
\Pi\rpcycl = \int \zetazrprp d\pi_{12} d\pi_{23}
\eeq
Because the total amount of triplets along the line-of-sight is not
distorted by the peculiar motions, the projected 3PCF $\Pi\rpcycl$ is
related to the 3PCF in real space $\zetarr$ :
\beq\label{3pcfproj2}
\Pi\rpcycl =\int \zeta(\sqrt{r_{p12}^2+y_{12}^2},\sqrt{r_{p23}^2+y_{23}^2},
\sqrt{r_{p31}^2+(y_{12}+y_{23})^2})dy_{12} dy_{23}\,.
\eeq

Similarly as for $\zetass$, we measure $\zetazrprp$ by
counting the numbers of triplets $DDD\rppicycl$, $DRR\rppicycl$,
$RDD\rppicycl$ and $RRR\rppicycl$ formed by galaxies and/or random points
with the projected separations $r_{p12}$, $r_{p23}$, and $r_{p31}$ and
radial separations $\pi_{12}$ and $\pi_{23}$.  We will use $r_p$, $u$
and $v$:
\beq\label{rpuv}
r_p = r_{p12}, \hskip1cm u={{r_{p23}}\over{r_{p12}}},\hskip1cm
v={{r_{p31}-r_{p23}}\over{r_{p12}}}\,.
\eeq
to quantify a triangle with $r_{p12}\le r_{p23}\le r_{p31}$ on the
projected plane. Equal logarithmic bins of intervals $\Delta \lg
r_p=\Delta \lg u=0.2$ are taken for $r_p$ and $u$, and equal linear
bins of $\Delta v=0.2$ for $v$. The same ranges of $u$ and $v$ are
used as for $\zetasu$, but $r_p$ is from $0.128\mpc$ to $4\mpc$ (7
bins). The radial separations $\pi_{12}$ and $\pi_{23}$ are from
$-25\mpc$ to $25\mpc$ with a bin size of $1\mpc$.  The projected
3PCF is estimated by summing up $\zeta_z(r_p, u,
v,\pi^i_{12},\pi^j_{23})$ at different radial bins
($\pi^i_{12},\pi^j_{23}$):
\beq\label{Pistat}
\Pirpu= \sum_{i,j} \zeta_z(r_p, u, v,\pi^i_{12},\pi^j_{23})
 \Delta\pi^i_{12} \Delta\pi^j_{23}
\eeq
and normalized as
\beq\label{qproj} 
\Qrpu ={\Pi(r_p,u,v) \over \wrp{12}\wrp{23}
+\wrp{23}\wrp{31} +\wrp{31}\wrp{12}} \,.  
\eeq 
where $\wrp{}$ is the projected two-point correlation function
(\cite{dp83} 1983; JMB98)
\beq\label{wrp}
\wrp{} = \sum_{i}\xi_z(r_p,\pi^i)\Delta \pi^i
\eeq
An interesting property of the projected 3PCF is that if the
three-point correlation function is of the hierarchical form, 
the normalized function $\Qrpu$ is not only a constant but also equal
to $Q$. Therefore the measurement of $\Qrpu$ can be used to test the
hierarchical form which was proposed mainly based on the analysis
of angular catalogs.

\subsection{N-body tests of the statistical methods}
To test the reliability of our statistical analysis and to demonstrate
the effects of the redshift distortion, the projection, and the fiber
collisions, we make use of the full simulation and the mock
catalogs. Because the mock samples are cluster-(under)weighted, we
have applied the same weighting to the full simulation to achieve a
proper comparison. To calculate the quantities in redshift space for
the full simulation, we assume that the third axis is along the
line-of-sight.

\begin{figure}
\epsscale{1.0} \plotone{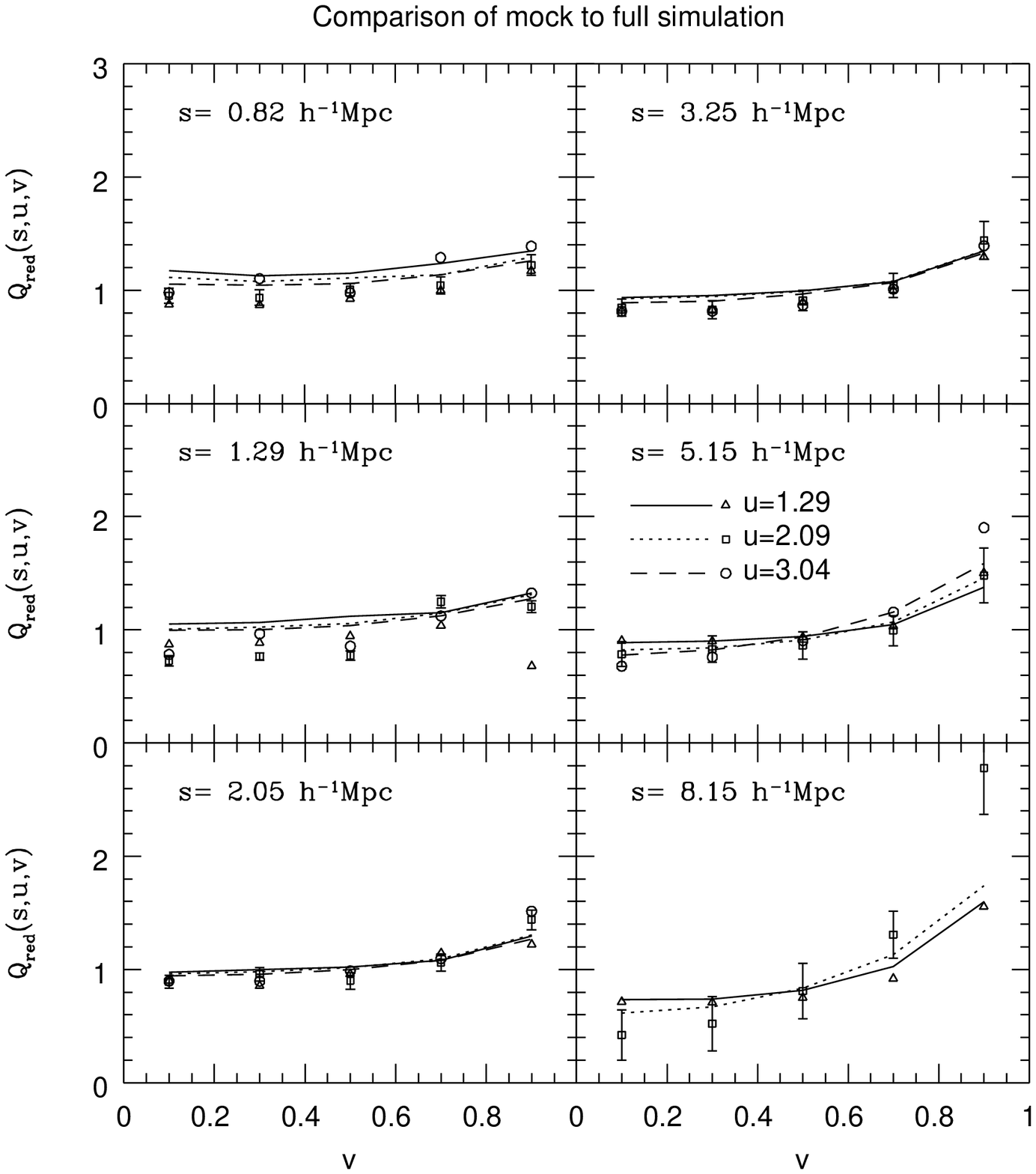}
\caption{
The normalized 3PCF in redshift space $\Qsu$ of the mock 
samples (symbols) and of the full simulation (lines). The error bars
are the $1\sigma$ standard deviation of the measurement for the 10
mock samples. For clarity, the error bars are plotted for $u=2$ only
but those for the other two values of $u$ are very similar.
}\label{fig1}\end{figure}

In Fig.~(\ref{fig1}) we compare  $\Qsu$ estimated with our
statistical method from the mock samples with the {\it true} value.
The latter is determined from the full simulation using the method of
\cite{jb97} (1997). On the scales from $1\mpc$ to $10\mpc$, the two estimated
quantities agree fairly well, indicating that the LCRS can yield an
unbiased estimate within the error bars of the 3PCF in redshift space.
The test is important, considering the fact that the LCRS is
essentially two-dimensional with one dimension in the direction of the
line-of-sight.

\begin{figure}
\epsscale{1.0} \plotone{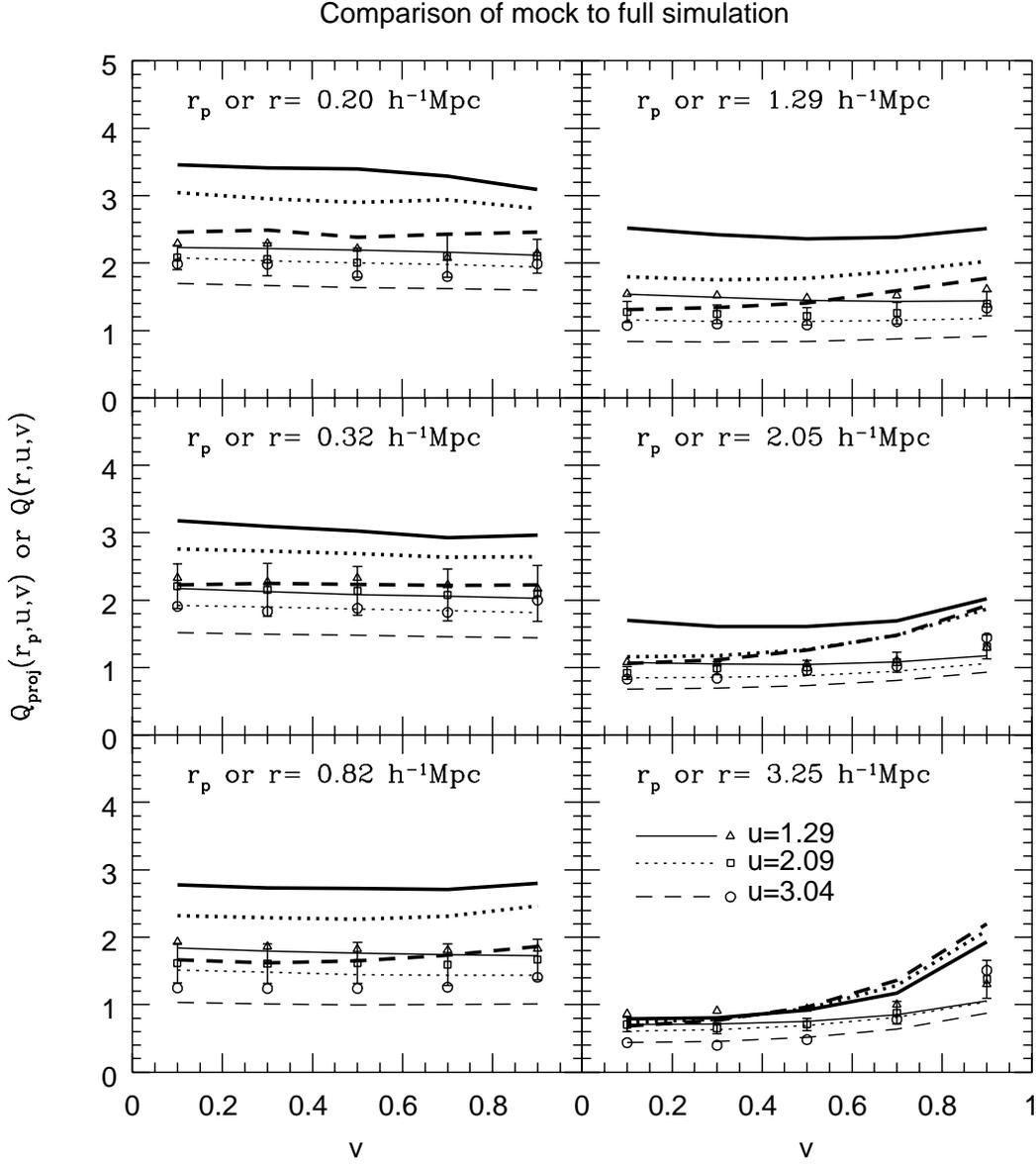}
\caption{
The normalized projected $\Qrpu$ of the mock samples
(symbols) and of the full simulation (thin lines). The latter is
computed through eq.~(\ref{3pcfproj2}) from the 3-D 3PCF $\Qru$ of the
full simulation which are also plotted with the thick lines. A
comparison of the symbols and the thin lines tests the statistical
methods, and a comparison of the thin lines and thick lines shows the
projection effect. As in Fig.~(1), the error bars are the $1\sigma$
standard deviation of the measurement for the 10 mock samples and are
plotted for $u=2$ only.  }\label{fig2}\end{figure}

Fig.~(\ref{fig2}) shows the projected $\Qrpu$ estimated from the mock
samples  with the
method described above. The true $\Qrpu$ can be
calculated from the {\it real space} 3PCF $\zetaru$ through
Eq.~(\ref{3pcfproj2}). We determine  $\zetaru$ for the full
simulation using the method of \cite{jb97} (1997) and calculate the
integral of Eq.(\ref{3pcfproj2}) by linearly interpolating the
estimated $\zetaru$. The two estimated quantities agree very well
within the error bars, indicating that our method can give a correct
estimate of the projected 3PCF.  The real space 3PCF $\Qru$ is also shown
in the figure by the thick lines.  It decreases with the scale $r$, as
noted previously (\cite{ms94} 1994; \cite{jb97} 1997). The reason for
the decrease is due to the fact that the slope of the power spectrum
is more negative on smaller scales (see Jing, in preparation, for a
detailed discussion). The consequence is that due to the averaging of
$\Qru$ on scales $r\ge r_p$ [Eq.(\ref{3pcfproj2})], the projected
$\Qrpu$ is also a decreasing function of $r_p$ but smaller than $\Qru$
for $r=r_p$ (compare thick and thin lines in the figure). Another
interesting point is that $\Qru$ is much higher than its counterpart
in redshift space $\Qsu$ on scales $\ls 10\mpc$. This result is
well-known since the redshift distortion smears out the dense clusters
and thus reduces the 3PCF on small scales (e.g. \cite{ms94} 1994;
\cite{m94} 1994).

We have also tested for the fiber collision effect of the LCRS sample
as we did for the 2PCF and PVD in JMB98. While both the 2PCF [$\xi(s)$
and $w(r_p)$] and the 3PCF [$\zetasu$ and $\Pirpu$] show some small dependence
on this effect, it cancels out completely, when we divide these two 
quantities by one another to form the normalized 3PCF $\Qsu$ and
$\Qrpu$. Therefore, the fiber collisions of the LCRS have little
effect on the normalized functions $\Qsu$ and $\Qrpu$.

In summary, these tests have convinced us that our method is suitable for
giving a stable measurement of the 3PCF from the LCRS with reliable
error estimates.

\subsection{The statistical results of the LCRS}
\begin{figure}
\epsscale{1.0} \plotone{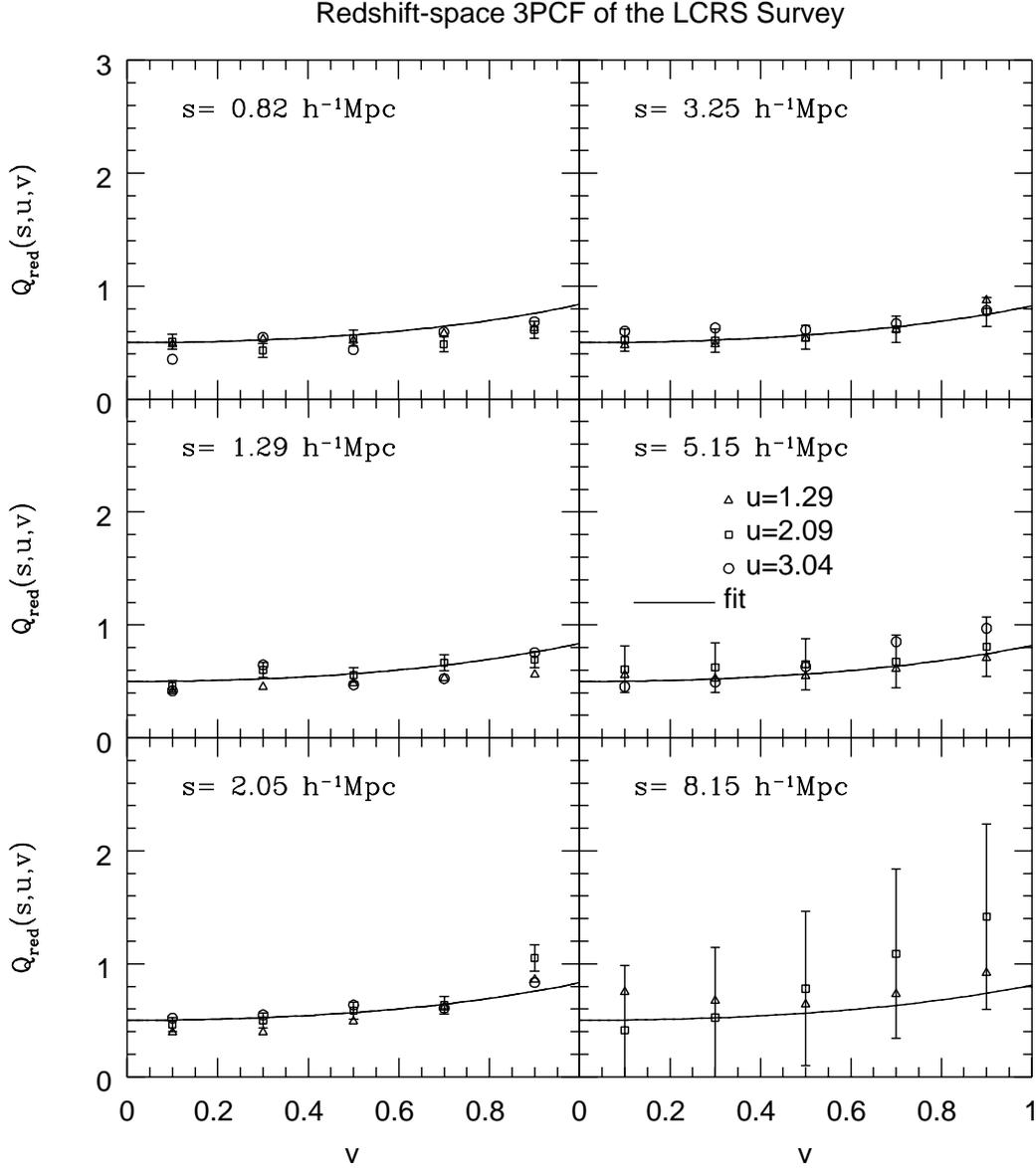}
\caption{ 
The normalized 3PCF in redshift space $\Qsu$ of the LCRS
survey (symbols). The errors are estimated by the bootstrap resampling
method. For clarity, the
error bars are plotted for $u=2$ only but those for the other two
values of $u$ are very similar.  The results are well fit by
$\Qsu=0.5\cdot 10^{[0.2+0.1({s\over s+1})^2]v^2}$ ($s$ is unit of $\mpc$)
 which are shown by the solid lines.}
\label{fig3}\end{figure}

\begin{figure}
\epsscale{1.0} \plotone{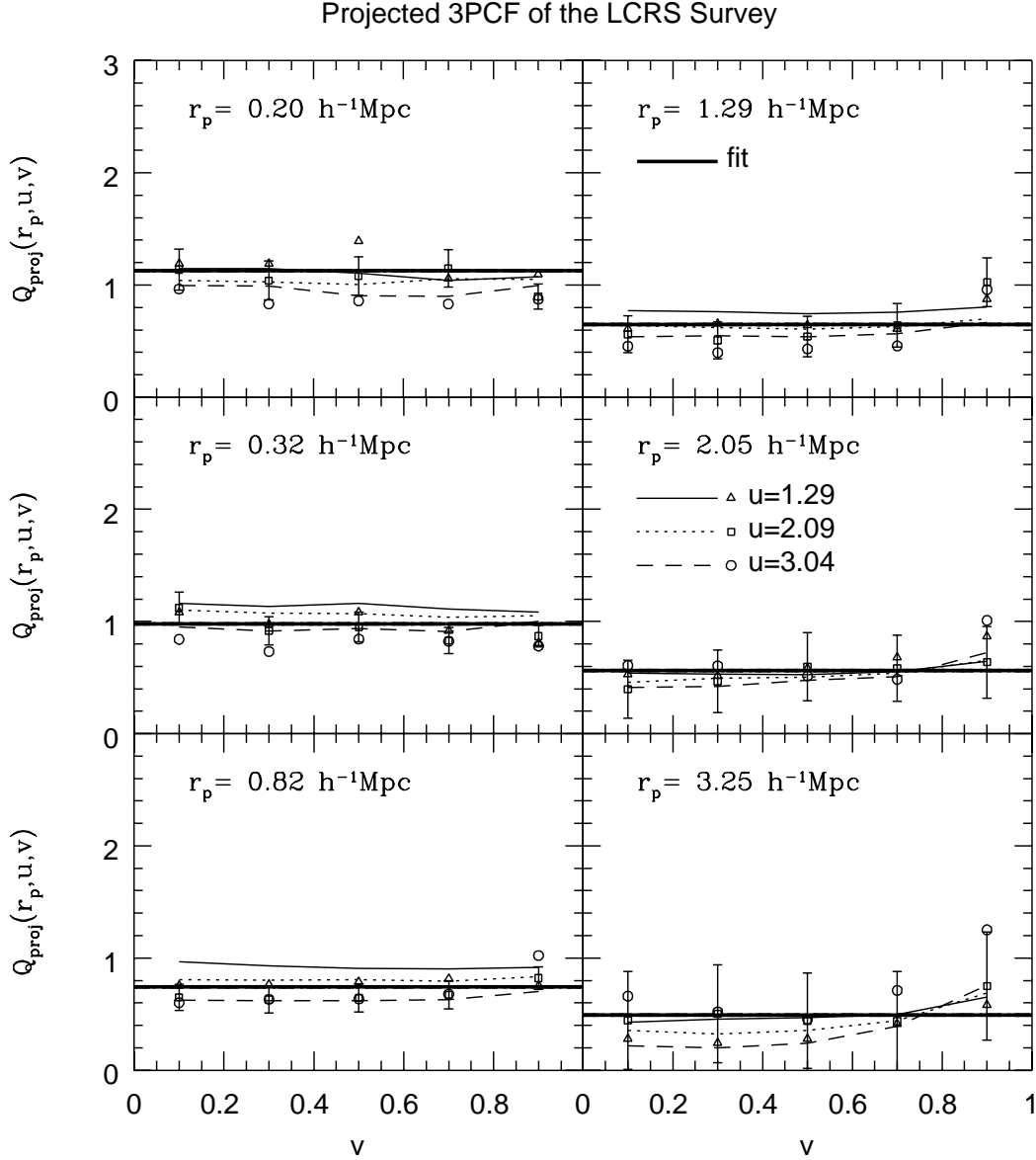}
\caption{
  The normalized projected 3PCF $\Qrpu$ of the LCRS survey
  (symbols). The errors are estimated by the bootstrap resampling
  method. For clarity, the error bars are plotted for $u=2$ only but
  those for the other two values of $u$ are very similar.  The results
  are well fit by $\Qrpu= 0.7(r_p)^{-0.3}$ ($r_p$ is unit of $\mpc$)
  which are shown by the thick solid lines.  The thin lines are a {\it
    half} of the mean projected 3PCF of the the mock samples, which
  seem to fit the observational data very well.  }
\label{fig4}\end{figure}

We present our results of the 3PCF in redshift space $\Qsu$ and of the
projected 3PCF $\Qrpu$ in Figures (\ref{fig3}) and
(\ref{fig4}) respectively for the Las Campanas Redshift Survey. The errors of the $Q$-values are the
bootstrap errors which are estimated with the approximate formula of
\cite{mjb92} (1992). As we can see from Fig.~(\ref{fig3}), the
3PCF obtained from redshift space is not changing very much with $s$
or $u$; it increases somewhat with $v$.  For small $v$, $Q_{red}$ is
approximately constant with a value of $ \sim 0.5$, but it increases
up to $\sim 1$ when $v\approx 1$.  Compared with the 3PCF in redshift
space, the projected one $\Qrpu$ [Figure~(\ref{fig4})] shows
quite similar dependences on triangle shape (i.e. $u$ and $v$), but
quite a different dependence on the triangle size (i.e. $s$ or $r_p$),
though the errors of the projected 3PCF are larger. Its value
decreases with $r_p$ from about $1.2$ at $r_p=0.2\mpc$ to $0.5$ at
$r_p$ about $2\mpc$. {\it Both this decrease with growing $r_p$ and
the weak increase with $v$ are in contrast to the hierarchical
assumption}. If the three-point correlation function in real space
were hierarchical, equation (\ref{3pcfproj2}) shows that the projected
one would also be hierarchical and equal to $Q$.  This behavior,
however, is qualitatively in agreement with the CDM model predictions,
and we will discuss this point in \S 4.

The hierarchical form (eq.1) does not seem to provide an adequate
description of our results of the 3PCF in redshift space $\Qsu$ or of
the projected 3PCF $\Qrpu$. We have looked for fitting formulae for
both quantities. Our results can be fitted quite well by
$\Qsu=0.5\cdot 10^{[0.2+0.1({s\over s+1})^2]v^2}$ and
$\Qrpu=0.7r_p^{-0.3}$ ($s$ and $r_p$ are in unit of $\mpc$), which are
the smooth lines on Figure (\ref{fig3}) and the thick lines on Figure
(\ref{fig4}). We have neglected the weak systematic dependence on $v$
of $\Qrpu$ at $r_p\gs 1\mpc$ in fitting this quantity since the
dependence is not statistically significant on the scales we
probed. We do not intend to give an error estimate for the
coefficients in the formulae, since the errors of the $Q$-values in
individual bins are likely non-Gaussian distributed and correlated
among different bins. These formulae are intended to give a simple,
but for most purposes accurate, approximation to our statistical
results.

Since the real space 3PCF $\zetaru$ possibly depends on $r$, $u$ and
$v$ in a complicated way, the inversion of eq.(\ref{3pcfproj2}) to get
$\zetaru$ from the projected $\Pirpu$ is certainly unstable.  We
have noted however that the projected $\Qrpu$ can be modeled very well
by {\it half} the value of the mock projected $Q_{proj}^{cdm}(r_p,u,v)$ [the
thin lines in Fig.~(\ref{fig4})], which means that
$\Qru=0.5Q^{cdm}(r,u,v)$, where $Q^{cdm}(r,u,v)$ is the real-space 3PCF
of the CDM model, is a good approximation to the
real-space 3PCF of LCRS galaxies on scales $\ls 3\mpc$. These $\Qru$
can be easily read out from the $Q^{cdm}(r,u,v)$ in Fig.~(\ref{fig2}).

\subsection{Discussion}
The high quality of the LCRS survey, by its CCD photometry, its
complete redshift information and its large size, has enabled us to
give a reliable determination of the 3PCF. It is the first time that
the three-point correlation functions both in redshift space and in
real (projected) space can be measured as a function of the triangle
size and shape, with only a fractional error in each individual bin,
for a wide range of triangle configurations. Although our statistical
results in  real space on scales $\ls 1\mpc$ are not far from the
hierarchical prediction (1) with $Q=1.3\pm 0.2$ (\cite{gp77} 1977;
GP77), the systematic changes of $\Qrpu$ with the triangle
configurations clearly point to a more elaborate model for $\Qru$. One
such model was already proposed in \S 3.3. Although our result of
$\Qsu$ generally agrees with the previous studies based on much
smaller redshift samples (see \S1), our measurement has much better
accuracy.

Recently there are concerns about the reliability of the high-order
correlation functions derived from photographic-plates based galaxy
catalogs. The skewness of the galaxies derived from the Automatic
Plate Machine (APM) angular catalog (\cite{mesl90} 1990) and
from the Edinburgh/Durham Southern Galaxy Catalogue (EDSGC,
\cite{hcm89} 1989) is significantly different even on small
scales $<1\mpc$ (\cite{g94} 1994; \cite{smn96} 1996), despite 
the fact that both catalogs are constructed from the same UK IIIa-J
Schmidt photographic plates and the latter is just a sub-sample ($\sim
1/4$) of the former. The difference seems to arise from their
different methods to digitalize the plates (\cite{sg98} 1998), and at
this point it is rather difficult to judge which method is
superior. It is also worth noting that the 3PCF results of GP77 
are somewhat sensitive to the
corrections they have applied to the Lick catalog.  Fortunately the LCRS
survey does not suffer from these uncertainties since its photometric
catalog was constructed from the CCD drift scans. Furthermore in our
analysis we did not have to apply any additional corrections except
for those well-documented by the survey team.

It is interesting to compare our results with the skewness $S_3(R)$
determined from the large angular catalogs, in particular
since the two studies based on the APM and EDSCG catalogs
have yielded rather discrepant results. The skewness is
related to the 3PCF through an integral as follows:
\bey\label{s3r}
S_3(R)&=&{\bar \zeta(R)\over {\bar\xi}^2(R)}\,\nonumber\\
\bar \zeta (R) &=&{1\over V^3}\int_{{\rm sphere}\, R} \,d{\mathbf r_1}
\,d{\mathbf r_2}\,d{\mathbf r_3} \zeta(r_{12},r_{23},r_{31})\,,\nonumber\\
\bar \xi (R) &=&{1\over V^2}\int_{{\rm sphere}\, R} \,d{\mathbf r_1}
\,d{\mathbf r_2} \xi(r_{12})\,,
\eey
where $V={4\pi\over 3} R^3$. It needs full information of $\Qru$ to
calculate the skewness [eq.~(\ref{s3r})], and we use the model
proposed in \S 3.3. The skewness $S_3(R)$ for the LCRS survey is then
about $4.5$ at $R=0.2\mpc$ and about $3.5$ at $R=1\mpc$, which seem in
agreement with the results of \cite{smn96} (1996) based on
the EDSGC catalog but significantly higher than the APM results of \cite{g94}
(1994) on the scales $< 1\mpc$. 
 
The cosmic virial theorem (CVT) has been widely used to measure the
mean density of the universe. If the 3PCF is hierarchical, the CVT can
be expressed in its simplified form relating the density parameter
$\Omega_0$, $\xi(r)$, $Q$, and the PVD $\sigma_{12}(r)$. Since our
results show that the 3PCF is not hierarchical, this relation becomes
much more complicated and it is necessary to work out the integration
over $\zetarr$ which might depend on $\zetarr$ on very small scales
$r\sim 0$. Since $\zetarr$ is a decreasing function of the triangle
size, previous studies which usually used the $Q$ value at $\sim
1\mpc$ might have overestimated the mean density.  The size of
galaxies, which are usually treated as point sources in the CVT
application, may also be important in the estimate of the mean
density, especially near $r\sim 0$ (\cite{p76} 1976; \cite{sj97} 1997).

We would like to remark here that the significant difference between
$Q_{ red}$ and $Q_{ proj}$ at small scales comes from peculiar
motions of galaxies.  This gives another possibility to estimate the
velocity dispersion of galaxies (\cite{m94} 1994).

\section{A case for model testing}
In this section we compare the 3PCF of the LCRS with model
predictions. \cite{jb97} (1997) and Jing (in preparation) have
recently studied the 3PCF $\Qru$ for a set of CDM models based on 
second-order perturbation theory and N-boby simulations. We found
that for fixed $u$ and $v$, $\Qru$ is a decreasing function of the
size $r$. In the strongly non-linear regime ($\xi(r)\gg 1$, $\Qru$
shows a very weak dependence on $u$ and $v$. In the weakly non-linear
and linear regimes, $\Qru$ increases significantly with $v$ for fixed
$r$ and $u$.  All these features are found in the projected $\Qrpu$ of
the LCRS galaxies. Therefore the statistical results found from the
LCRS survey are all qualitatively consistent with the mass 3PCF based
on N-body simulations of cosmological models.

\begin{figure}
\epsscale{1.0} \plotone{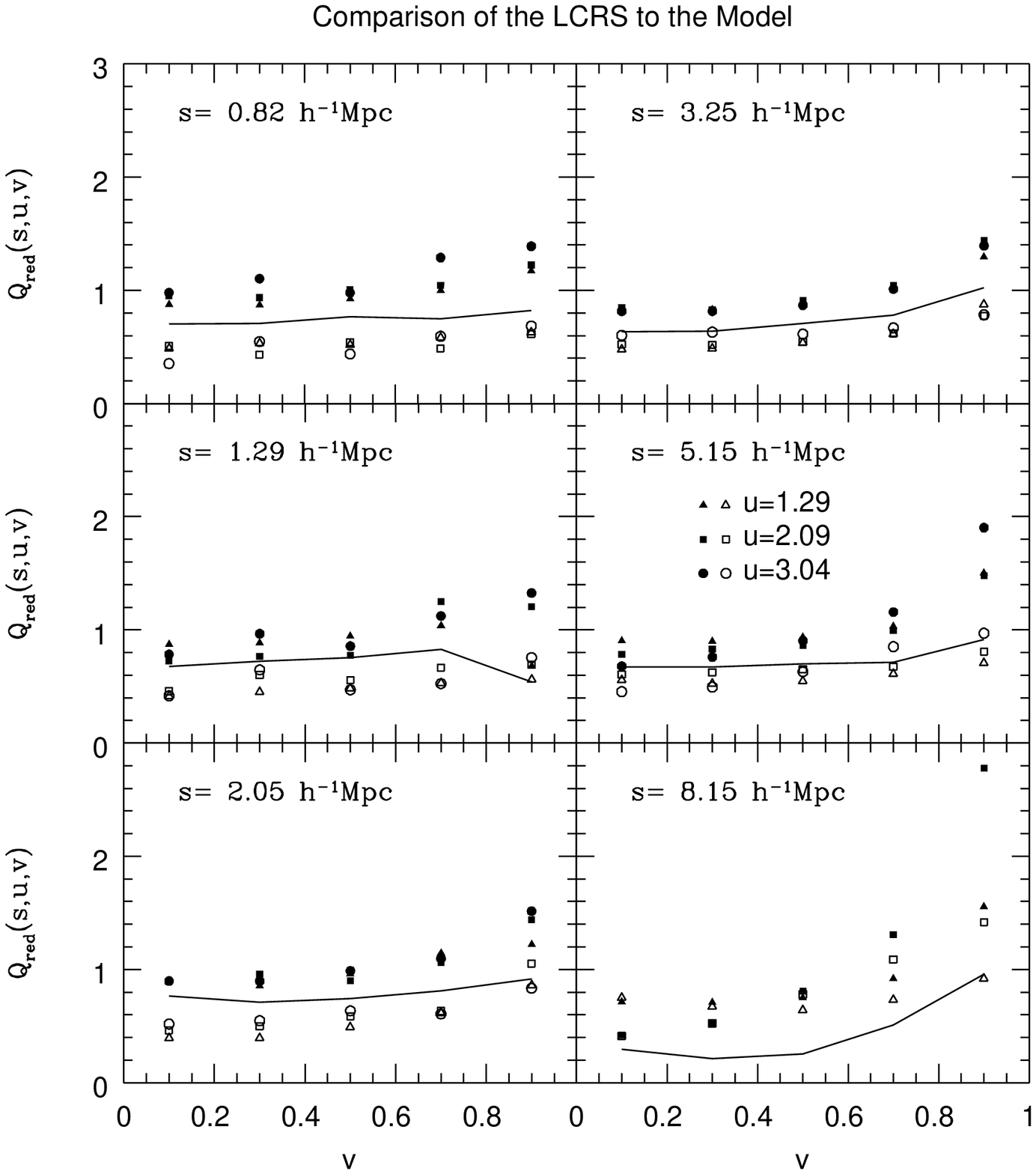}
\caption{ 
The normalized 3PCF in redshift space $\Qsu$ 
of the LCRS
survey (open symbols) compared with the mean value of the ten mock samples
(solid symbols). The thick solid lines represent the lowest of the ten
mock $\Qsu$ values in each $u=1.29$ bin, which can be regarded as the lower
model limits at $\sim 90\%$ significance level.
}\label{fig5}\end{figure}

\begin{figure}
\epsscale{1.0} \plotone{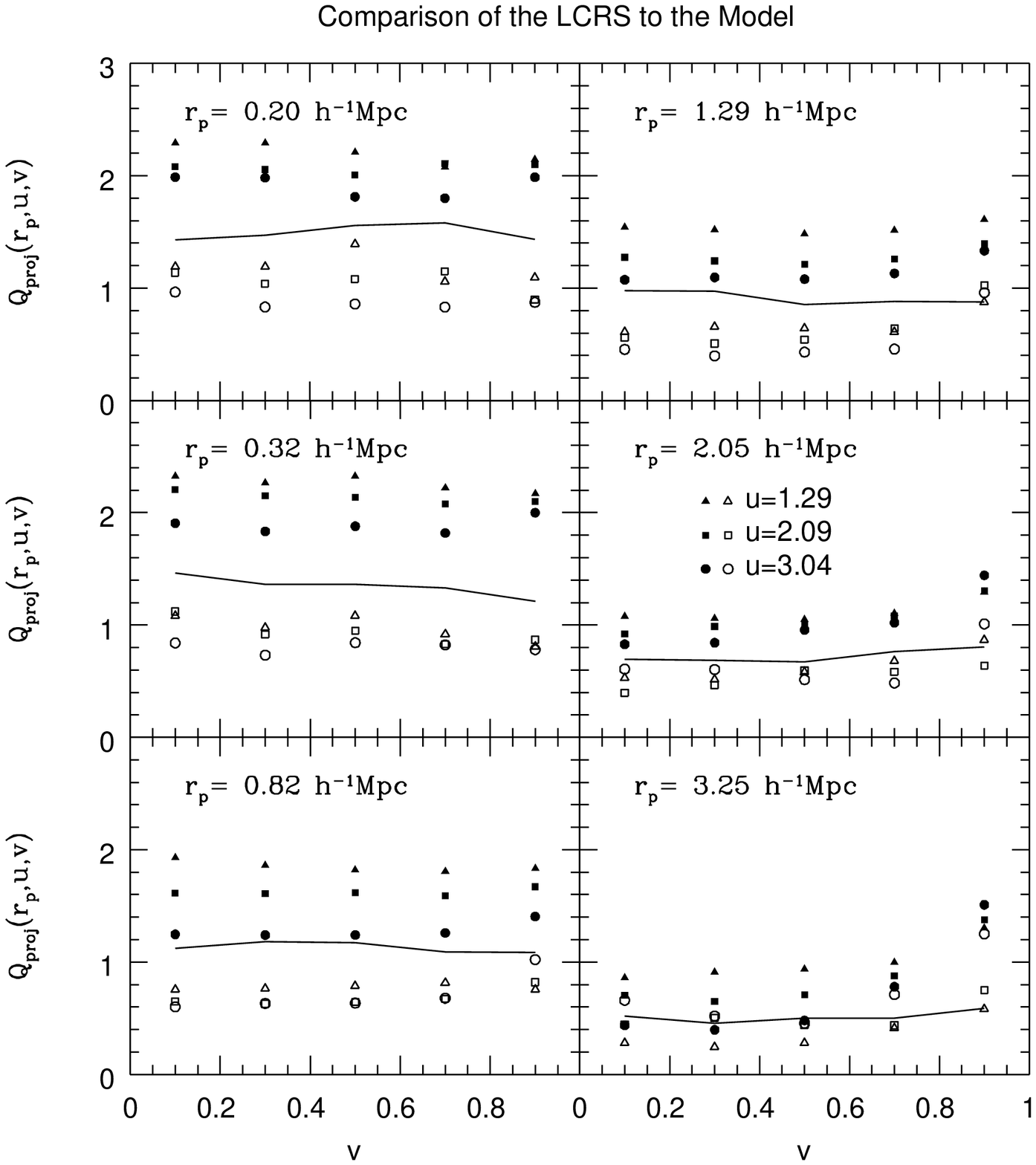}
\caption{ The normalized projected 3PCF of the LCRS survey (open
symbols) compared with the mean value of the ten mock samples (solid
symbols). The thick solid lines represent the lowest of the ten mock
$\Qsu$ values in each $u=1.29$ bin, which can be regarded as the lower model
limits at $\sim 90\%$ significance level.  }\label{fig6}\end{figure}

As an example to quantitatively test models with the 3PCF, we compare
the 3PCFs of the LCRS galaxies with the results of the mock samples in
Figs.~(\ref{fig5}) and (\ref{fig6}).  Only in this way could the
redshift distortion and the projection effects be accounted for
properly.  From the figures we find that the qualitative features,
i.e. the dependence on $v$ for fixed $s$ or $r_p$, and $u$, the
decrease of $Q$ with increasing values of $s$ or $r_p$ are reproduced
quite well in the mock samples. The values of the data set, however,
are consistently lower than the mean model predictions by a factor
$\sim 2$. Since the $Q$-values in each bin of the ten mock samples are
not Gaussian distributed (skewed to high values), it is not meaningful
to use the standard deviation to quantify the statistical
significance. Instead we pick up the lowest of the ten mock $Q_{red}$
or $Q_{proj}$ values in each bin and compare it with the observed
results.  The thick lines in Figs.~(\ref{fig5}) and (\ref{fig6})
correspond to these lowest values for $u=1.29$, which should be
compared with the open triangles. These lines are still higher than
(in most bins) or at least as high as (in a few bins) the
observational values, which indicates that the observational values
are lower than this model's predictions at a confidence level $\gs
90\%$. The underlying model for the mock sample is a CDM universe with
$\Omega_0=0.2$ and $\lambda_0=0.8$ and with clusters underweighted
(see \S 2). We have computed the 3PCF for this universe {\it without}
cluster weighting, and found the cluster weighting, like the fiber
collision effect, does not change much the value of $\Qru$. Thus, even
if this model fits the 2PCF and the PVD of the LCRS galaxies quite
well, it seems not really adequate to describe the clustering of
galaxies when the 3PCF is considered, unless the observed 3PCF is
biased low by the cosmic variance [at $\ls 10\%$ probability] .  This
might indicate that the gravitational interaction alone is not
sufficient to describe the clustering of galaxies, and physical
processes of gas and radiation hydrodynamics connected with galaxy
formation must be taken into account. A positive bias, i.e. a biasing
parameter $b>1$, can reduce the 3PCF. But, perhaps this conclusion
goes too far -- in fact, a slightly higher shape parameter $\Gamma$
will give a better fit, because $Q$ becomes smaller if $\Gamma$ is
increased. It appears that a model with a new set of parameters is to
be sought and the 3PCF determined here should provide to such new
models a test in addition to the 2PCF and the PVD. 

\section{Conclusion}
The result is clear, and the conclusions are straightforward: We have
succeeded in measuring the 3PCF from the LCRS. This is the first time
that the three-point correlation function of galaxies has been
measured accurately from a redshift survey. Both the 3PCF in redshift
space and the projected 3PCF have been measured as a function of the
triangle size and shape with only a fractional uncertainty in each
individual bin. Various tests have been carried out to assure that the
measurement is stable and that the errors are estimated reliably.  Our
results indicate that the 3PCFs both in redshift space and in real
space have small but significant deviations from the well-known
hierarchical form.  The 3PCF in the redshift space can be fitted by
$\Qsu=0.5\cdot 10^{[0.2+0.1({s\over s+1})^2]v^2}$ for $0.8<s_{12}<8
\mpc$ and $s_{31}<16\mpc$, and the projected 3PCF by
$\Qrpu=0.7r_p^{-0.3}$ for $0.2<r_{p12}<3\mpc$ and $r_{p31}<6\mpc$ ($s$
and $r_p$ are in unit of $\mpc$).  Although it might be not unique to
get 3-D real space $\Qru$ from the measured projected function
$\Qrpu$, we found that a half of the predicted $\Qru$ of the CDM model
considered in this paper provides a good description of the LCRS data.

The three-point correlation function gives an additional statistical
tool to constrain cosmogonic models. The general dependence of the
3PCF on triangle shape and size is in qualitative agreement with the
CDM cosmogonic models. Quantitatively the 3PCF of the models may
depend on the biasing parameter and the shape of the power spectrum,
in addition to other model parameters. Taking our result together with
the constraints imposed by the two-point correlation function and the
pairwise velocity dispersion of galaxies also obtained from the LCRS,
we find that we have difficulties to produce a {\it simple} model that
meets all constraints perfectly.  Among the CDM models considered, the
flat model with $\Omega = 0.2$ meets the 2PCF and PVD constraints, but
gives higher values for the 3PCF than  observed.  This may indicate
that more sophisticated bias models or a more sophisticated
combination of model parameters must be considered.

\acknowledgments 

We are grateful to Yasushi Suto for helpful
discussions, and for the hospitality extended to us at the physics
department of Tokyo university.  G. B. thanks the Yamada foundation
for support during his stay at RESCEU. J.Y.P. gratefully acknowledges
the receipt of a JSPS postdoctoral fellowship. Support by SFB375 is
also acknowledged.  The simulations were carried out on VPP/16R and
VX/4R at the Astronomical Data Analysis Center of the National
Astronomical Observatory, Japan.

\end{document}